# Changing redshifts caused by a changing expansion rate of the universe.

(October 11, 2024)

Nico Roos, Eric Sluimer and Bert van den Broek

*Affiliation: Leiden Observatory, PO Box 9513, NL-2300 RA Leiden, The Netherlands*

*(Revised version: August 12, 2026)*

## ABSTRACT

With the next generation of big telescopes such as the ELT and SKA it might become possible to measure changes in the expansion rate of the Universe in real time by measuring the change of the redshifts of a large number of galaxies over a period of the order of 10 years. This phenomenon, known as 'redshift drift,' will provide a crucial direct test of cosmological models. The change in redshift is readily explained using the concept of conformal time which is the comoving distance of a galaxy in lightyears. We emphasize that the redshift drift is directly proportional to the average change in the cosmic expansion rate between the time of a galaxy's light emission and its absorption. This phenomenon is illustrated within the framework of the concordance model, the Lambda-CDM model of the universe.

## I. INTRODUCTION

The cosmological expansion causes the redshift of photons travelling through the universe. This phenomenon allows us to estimate the basic expansion parameters at the present cosmological time $t_0$: the Hubble constant $H_0$ and the deceleration parameter $q_0$. [1-3] These parameters can then be used to infer the parameters of a particular cosmological model. Before the turn of the millennium it was generally assumed that the universe was decelerating. However studies of the supernovae in distant galaxies indicated that, over the last billion years, the universe has transitioned from deceleration to acceleration.[4] This observation was interpreted as evidence for a cosmological constant (Dark Energy, or DE) exerting a negative pressure.

Observations of the Hubble flow of galaxies, the cosmic background radiation (CBR) and baryon acoustic oscillations (BAO's) can be interpreted in the flat Cold Dark Matter Model with a cosmological constant (Λ), the $\Lambda CDM$ model. However, the dominant energy components of this model, CDM and Λ, lack a solid physical basis. Furthermore, recent observations suggest that the cosmological constant might not actually be constant. [5,6] New observations are therefore essential to better understand the expansion history of the universe.

The next generation of very large telescopes is likely to play a fundamental role in determining the recent acceleration history of the universe by studying the evolution of the redshift of galaxies in real time.[7-13] This so called "redshift drift" is the result of the changing expansion rate of the universe. In a "coasting" universe, where the expansion rate remains constant, the redshift of galaxies would not change. In a decelerating universe galaxies would appear to become bluer while in an accelerating universe, they would become redder.

In this paper we use the concept of conformal time to demonstrate how the redshift of a galaxy changes when observed at different cosmological times. We emphasize that the change in redshift directly results from the average change in the cosmological expansion rate between the times of light emission and absorption. We begin by introducing the standard cosmological model relation between comoving distance and conformal time. In section III we introduce the concept of conformal time and in IV we show that a graph of the logarithm of the scale factor (or expansion factor) versus the logarithm of the conformal time illustrates how redshift drift relates to the gradient of the scale factor in this diagram. The diagram also shows how lines of constant redshift behave in a conformal time – comoving space diagram, which we briefly discuss in section V. In section VI, we discuss the redshift drift in more detail.

## II. THE STANDARD COSMOLOGICAL MODEL

In the standard isotropic homogeneous Friedmann-Lemaitre-Robertson-Walker (FLRW) cosmological model the radial part of the metric is given by

$$ds^2 = c^2dt^2 - dr^2 = c^2dt^2 - a(t)^2d\chi^2\,, \tag{1}$$

where $a(t)$ is the scale factor of the universe as a function of cosmological time, $r$ is the proper distance and $\chi$ is the comoving distance. The scale factor is determined by the cosmological model via the Einstein equations. For illustration we adopt a flat $\Lambda CDM$ model which includes pressureless baryonic plus Cold Dark Matter (CDM) and Dark Energy (DE) represented by the cosmological constant Λ. While this model also contains a radiation component, we omit it here since it does not significantly affects the universe's dynamics during the later evolutionary stages of evolution that we are considering.

The first and second derivatives of the scale factor are given by the Friedmann equations

$$\frac{da}{dt} = \dot{a} = H_0\left[\frac{\Omega_m}{a} + \Omega_\Lambda a^2\right]^{1/2}, \tag{2}$$

and

$$\ddot{a} = H_0^2\left[-\frac{\Omega_m}{2a^2} + \Omega_\Lambda a\right]. \tag{3}$$

Here the Hubble parameter is given by $H = \dot{a}/a$. $H_0$ is the Hubble parameter at the present epoch $t_0$ (normalizing $a$ as usual by setting $a(t_0) = 1$). $\Omega_m$ and $\Omega_\Lambda$ represent the density parameters for matter (baryonic and dark) and dark energy, respectively, normalized to the critical density $\varrho_{crit} = 3/8\pi G$. The model is flat, implying $\Omega_m + \Omega_\Lambda = 1$. Observations of the Cosmic Microwave Background Radiation (CMBR) and Supernova observations are in good agreement with $\Omega_m = 0.3$ and $\Omega_\Lambda = 0.7$. These values define the 'standard' model we adopt in this paper.

The first Friedmann equation can be generalized by using the energy conservation equation in combination with the equation of state for the relation between the pressure and the mass density ($p = w\rho$) for each component:

$$\dot{a} = H_0 a \sqrt{\sum_i \Omega_i a^{-3(1+w_i)}}\,. \tag{4}$$

DM is assumed to be pressureless ($w = 0$), while DE is usually characterized $w_{\mathrm{DE}} = -1$ corresponding to a cosmological constant. However, recent observations suggest that $w_{\mathrm{DE}}$ might vary with the scale factor. The redshift is given by the ratio of the scale factor at the moment of measurement (absorption) to that at emission:

$$1 + z = \frac{a_{abs}}{a_{em}}. \tag{5}$$

We can transform the first Friedmann equation (Eq. 2) into a simple form by using the substitution variable $\theta$ defined by $\tan\theta = (a/a_{m\Lambda})^{-3/2}$ where $a_{m\Lambda} = (\Omega_m/\Omega_\Lambda)^{1/3}$ and $k = \frac{3}{2} H_0\, \Omega_\Lambda^{1/2}$:

$$kdt = -\frac{1}{\sin\theta} d\theta\,, \tag{6}$$

which has the solution

$$kt = \ln(\cot\frac{\theta}{2}) = \ln\sqrt{\frac{1+\cos\theta}{1-\cos\theta}}\,. \tag{7}$$

The definition of $\tan\theta$ implies $\cos\theta = (1 + a_{m\Lambda} a^{-3})^{-1/2}$, giving us

$$t = k^{-1} \ln\left[\left(\frac{a}{a_{m\Lambda}}\right)^{3/2} + \sqrt{1 + \left(\frac{a}{a_{m\Lambda}}\right)^{3}}\right]. \tag{8}$$

The present cosmological time is then

$$t_0 H_0 = \frac{2}{3}\, \Omega_\Lambda^{-1/2} \ln\left[\frac{\sqrt{\Omega_\Lambda} + 1}{\sqrt{\Omega_\mathrm{m}}}\right], \tag{9}$$

and the scale factor can be expressed in the simple form

$$a(t) = a_{m\Lambda} \left[\frac{e^{2kt} - 1}{2e^{kt}}\right]^{2/3} = a_{m\Lambda}\, sinh^{2/3}(kt)\,, \tag{10}$$

With the Hubble parameter given by

$$H(t) = \frac{2}{3} k \left[\frac{e^{2kt} + 1}{e^{2kt} - 1}\right] = \frac{2}{3} k \frac{\cosh(kt)}{\sinh(kt)}\,. \tag{11}$$

While this model has become quite popular, it raises several fundamental questions, for instance what is DM and what is the physical interpretation of DE. If DE is interpreted as vacuum energy with $w = -1$ it is expected to have an energy density which is many orders of magnitude larger than the observed value. Additionally, there are two (independent) coincidence problems: 1) why is the energy density of DE about equal to the matter energy at the present epoch, and 2) why is the product of the age of the universe and the Hubble parameter almost equal to 1 ($t_0 H_0 = 0.964$ for our standard model), or even equal to 1 within observational uncertainties. [14] The latter problem might suggest that the universe has entered a "coasting" phase with $a \propto t$ (implying $tH(t) = 1$), during which the acceleration is zero and redshifts do not change over cosmological time.

Recent observations appear to indicate that $w_{\mathrm{DE}}$ might not be constant,, potentially following a relation like $w_{\mathrm{DE}} = w_0 + w_a(1-a)$, with $w_0 \sim -1, w_a \sim -0.7$ (see also section VI). [5,6] These unresolved questions and new observations highlight the necessity for further research to improve our understanding of the universe's dynamic history.

## III. CONFORMAL TIME AND COMOVING DISTANCE

We aim to study the change of the redshift of a galaxy due to the changing scale of the universe. Assuming that the galaxy has no significant peculiar motion, its comoving distance remains constant. From the metric we see that the light emitted from the galaxy in our direction will have a proper velocity

$$\frac{dr}{dt} = rH(r) - a(t)\frac{d\chi}{dt} = rH(r) - c\,. \tag{12}$$

The last equality follows from the metric by noting that $ds = 0$ for photons, meaning that the photon's proper velocity 'locally' always equals $c$. The conformal time $d\tau$ is defined as the time it takes for a photon to travel a comoving distance $d\chi$

$$d\tau = \frac{d\chi}{c} = \frac{1}{a(t)}dt\,. \tag{13}$$

The comoving distance light travels from the time of emission ($t_{em}$) from the galaxy to the time of absorption ($t_{abs}$) is

$$\chi = c\int_{t_{em}}^{t_{abs}} \frac{1}{a(t)}dt = c\int_{a_{em}}^{a_{abs}} \frac{1}{\dot{a}a}da\,. \tag{14}$$

From the final equality, we can infer that the redshift of a galaxy remains unchanged when the universe expands at a constant rate. In this case, we find that $\chi \propto \ln(1+z)$ which implies that z must remain constant.

## IV. SCALE FACTOR VERSUS CONFORMAL TIME

It is instructive to examen the redshift drift of a galaxy over cosmological time (or as a function of the scale factor) using the  relation between cosmological time (or scale factor) and conformal time under the condition that the conformal time (or comoving distance) remains constant. It is instructive to analyse this in conjunction with a graph of the logarithm of the scale factor as a function of conformal time (in units of $H_0^{-1}$) which is the comoving distance divided by $c$.

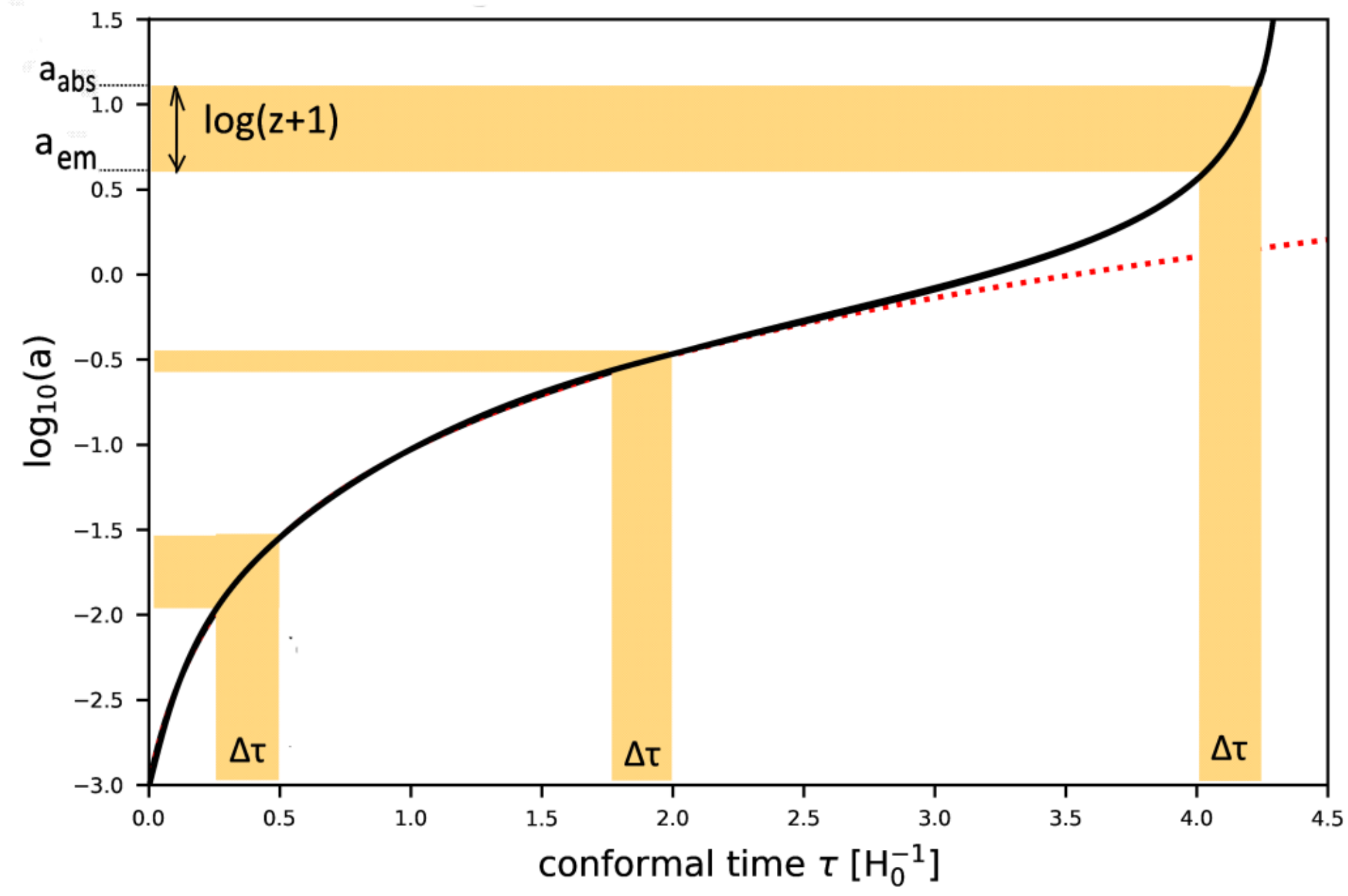

Figure 1. Logarithm of the scale factor for the standard $\Lambda CDM$ model as a function of conformal time. The vertical columns represent a galaxy at a comoving distance equal to $c\Delta\tau$. The redshift of the galaxy is given by $log_{10}(1+z) = log_{10}\, a_{abs} - log_{10}\, a_{em}$. The graph shows that the redshift will initially decrease and then increases over time. The redshift drift is determined by the gradient along the curve. As the cosmological constant Λ becomes dominant, the conformal time converges to a finite value. This convergence leads to an increase in redshift over time, in contrast to models without acceleration such as those with $\Lambda = 0$ (represented by the red dotted line).

The redshift of a galaxy with comoving distance $c\Delta\tau$ has a redshift given by $\log_{10}(1+z) = \log_{10} a_{abs} - \log_{10} a_{em}$. In Fig.1 we observe that when $\Delta\tau$ is kept constant the redshift initially decreases and the increases with cosmological time. The redshift drift follows the gradient of in the curve in the figure:

$$\frac{d\ln a}{d\tau} = \dot{a}\,. \tag{15}$$

This statement will be confirmed in the section VI (see Eq. 24). On the left sided of the curve, the gradient of the curve decreases, so $\dot{a}_{em} > \dot{a}_{abs}$ , resulting in a decreasing redshift is. Conversely, on the right side of the curve, as the expansion accelerates, the redshift begins to increase. At the inflection point, where $\ddot{a} = 0$, the redshift will remain unchanged (when $a_{em}$ and $a_{abs}$ are not to far from the inflection point), as we observed at the end of Section III. Figure 1 also illustrates that the comoving distance ($c\Delta\tau$) along lines of constant redshift first increases and then decreases with cosmic time (see Fig. 2 in Section V )

## V. CONFORMAL TIME – COMOVING DISTANCES DIAGRAM

It is instructive to examine the lines of constant redshift within a conformal time – comoving distance diagram. In such a diagram the Particle Horizon, the Past Light Cone and the Event Horizon appear as straight lines since they are determined by the comoving distance light has travelled in a certain amount of cosmic time between emission and observation. These are defined as follows:
Particle horizon (PH): The maximum comoving distance that light can have travelled since the Big Bang:

$$\chi_{PH} = c \int_{t=0}^{t} \frac{1}{a(t')} dt' = c\,\tau\,. \tag{16}$$

Past Light Cone (PLC) :The comoving distance that light has travelled since cosmic time $t$. All photons that we currently receive follow this trajectory:

$$\chi_{PLC} = c \int_{t}^{t_0} \frac{1}{a(t')} dt' = c \int_{0}^{t_0} \frac{1}{a(t')} dt' - c \int_{0}^{t} \frac{1}{a(t')} dt' = c\tau_0 - c\tau\,. \tag{17}$$

Event Horizon (EH): the comoving distance light can travel rom time $t$ to $t = \infty$. The EH exists only if the following integral converges:

$$\chi_{EH} = c \int_{t}^{\infty} \frac{1}{a(t')} dt'. \tag{18}$$

Accelerating models, such as models with positive Λ, have a finite conformal time as $t$ approaches infinity resulting in an Event Horizon (see section V). The acceleration in these models causes the redshift to increase over time. In contrast, decelerating models, such as those with (dark) matter, but $\Lambda = 0$ (the red dotted line in Fig.1), have $\dot{a}_{em} \geq \dot{a}_{abs}$ leading to a consistently decreasing redshift. In these models, conformal time does not converge to a finite value as cosmological time approaches infinity, indicating the absence of an Event Horizon.

Figure 2 illustrates the Particle Horizon (PH), the Event Horizon (EH), and the Past Light Cone (PLC). The thick black line in this picture represents the present age of the universe ($a = 1, t = t_0$). In the standard $\Lambda CDM$ model, the conformal time is approximately (adopting $H_0 = 70\ km\ s^{-1} Mpc^{-1}$):

$$\tau_0 H_0 = H_0 \int_0^{t_0} \frac{1}{a(t')} dt' \approx 3.3 \quad => \quad \tau_0 \approx 46.2\ 10^9 yr \tag{19}$$

As $t$ approaches infinity, $\tau_0 H_0$ converges to 4.4.

The blue region represents the area within the Hubble radius given by $v_H = c = Ha\chi$. The comoving distance tot he Hubble radius is given by

$$\chi = \frac{c}{aH} = \frac{c}{\dot{a}} = \frac{c}{H_0 \sqrt{\frac{0.3}{a} + 0.7a^2}}\,. \tag{20}$$

As time progresses the PLC shifts parallel upwards along with the horizontal black line. In contrast, the PH and the EH remain fixed over time. The comoving distance to the PH represents the distance that light (or relativistic particles) could have travelled since the Big Bang. The white region under the PH is the part of the universe which has become visible to us since the BB. The gray region on the right side of the PH and between the PLC and the EH is the part of the universe that is not yet visible to us. Photons emitted from outside the EH (the dark region) will never reach us. Galaxies move upward along worldlines of constant comoving distance. This implies that their redshifts will initially decrease and then increase, unlike the redshift of the Cosmic Microwave Background Radiation, which was emitted at a fixed cosmological time rather than a fixed comoving distance. The photons in the CMB that we receive at the present time ($t_0 \approx 13.5\ Gyr$) were emitted at the epoch of recombination, about 380.000 $yr$ ($z \approx 1100$) after the BB at a comoving look back distance of approximately $\chi{\sim}3.3cH_0^{-1}$ ($45\ Gly$ ). This distance is sometimes called "radius of the (present) visible universe" (the white triangle under the PH).

A galaxy at a comoving distance of $\chi{\sim}2cH_0^{-1}$ moves upward in time along the thin black vertical line that intersects the PLC at conformal time $\tau{\sim}1.4\ H_0^{-1}$. We might observe it as a quasar or a "little red dot" found by the James Webb Space Telescope at a redshift of $\sim 6-8$. As the intersection point of the galaxy's comoving position and the PLC moves upward from $\tau = 0$ the $z-$value of the intersection point descends from high values to a minimum and then increases again. The photons we observe today were emitted at $\tau{\sim}1.3\ H_0^{-1}$, that is before the galaxy has crossed the PH at $\tau_{PH}{\sim}2\ H_0^{-1}$. Photons that were emitted at that time will reach us at $\tau = 2\tau_{PH}{\sim}4\ H_0^{-1}$ with redshift $\sim 10$. When the galaxy approaches the EH at $\tau{\sim}2.5\ H_0^{-1}$ its redshift will go to infinity. Photons emitted after that time will never reach us.

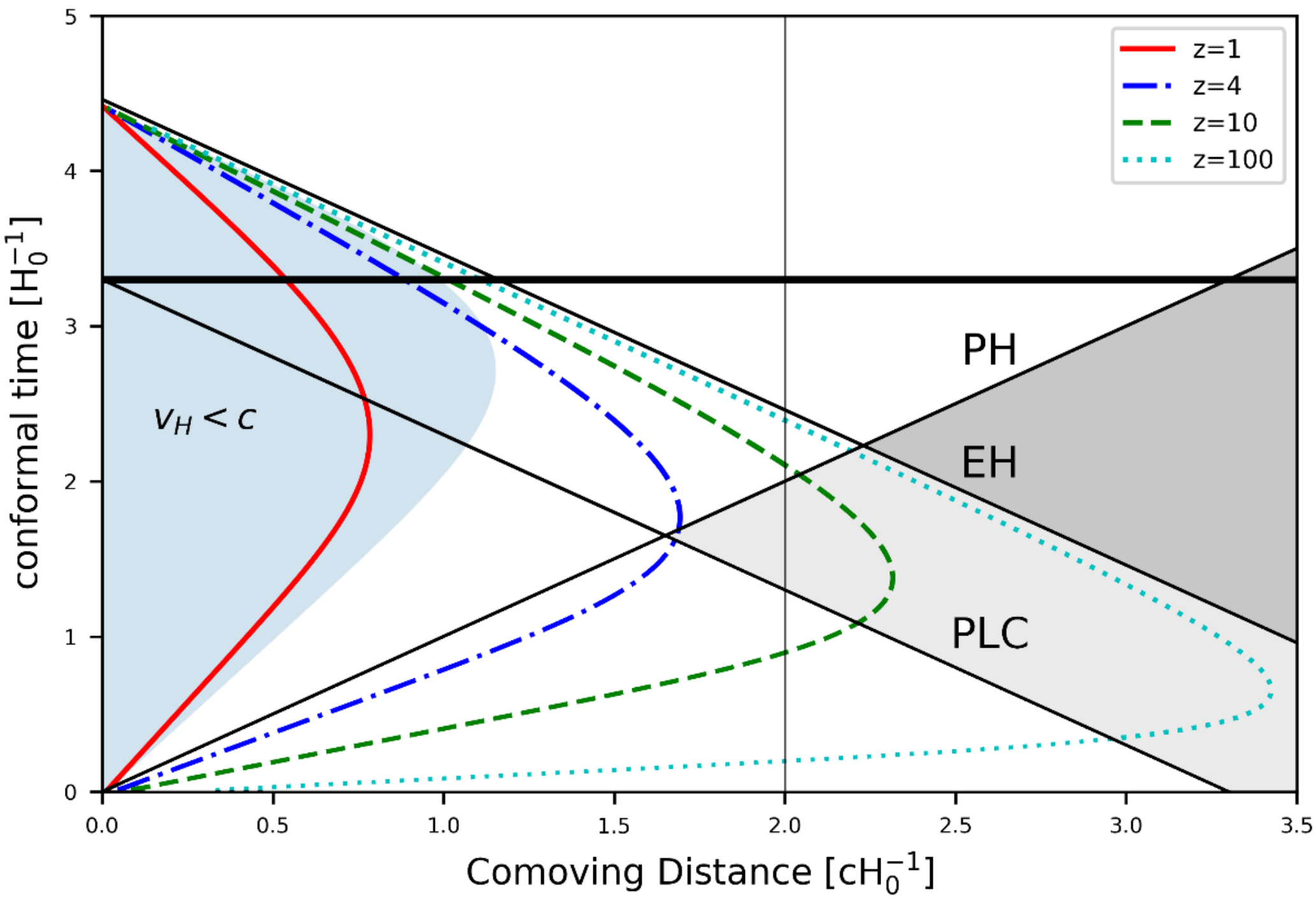

Figure 2. Conformal Time - Comoving distance diagram (see text)

# VI. REDSHIFT DRIFT AND CHANGING COSMIC EXPANSION

To derive the formula for the change of redshift of a galaxy we begin with the condition that the galaxy maintains a fixed comoving distance

$$\Delta\chi = c\Delta\tau = c\int_{t_{em}}^{t_{abs}} \frac{1}{a} dt = c\int_{a_{em}}^{a_{abs}} \frac{1}{\dot{a}a} da = \text{constant}\,, \tag{21}$$

where $a_{em} = a(t_{em})$ and $a_{abs} = a(t_{abs})$. According to the Leibniz integral rule, this can be expressed as

$$\frac{d\Delta\tau}{dt_{abs}} = \frac{d}{dt_{abs}}\int_{a_{em}}^{a_{abs}} \frac{1}{\dot{a}a} da = \\ = \frac{1}{a_{abs}\dot{a}(t_{abs})}\frac{da_{abs}}{dt_{abs}} - \frac{1}{a_{em}\dot{a}(t_{em})}\frac{da_{em}}{dt_{abs}} = 0\,. \tag{22}$$

From this we obtain

$$\frac{dt_{abs}}{dt_{em}} = \frac{a_{abs}}{a_{em}} = (1+z)\,. \tag{23}$$

Next, differentiate $(1+z) = a(t_{abs})/a(t_{em})$ with respect to $t_{abs}$ and combine it with the previous equation to find

$$\frac{dz}{dt_{abs}} = (1+z)H_{abs} - H_{em} = \frac{\dot{a}_{abs-}\dot{a}_{em}}{a_{em}}\,. \tag{24}$$

It shows that the redshift drift is due to the change in expansion velocity. A decreasing expansion velocity results in a decreasing redshift, while an increasing expansion velocity leads to an increasing redshift.

We can make a more general statement about the redshift drift for a universe in which the expansion velocity is changing arbitrarily. Let $Q(t_{abs}, t_{em})$ be the mean value of the acceleration of the expansion over the interval $\Delta t = t_{abs} - t_{em}$. This can be expressed as

$$Q(t_{abs}, t_{em}) = \frac{1}{\Delta t}\int_{t_{em}}^{t_{abs}} \frac{d}{dt}\dot{a}\, dt = \frac{\dot{a}_{abs-}\dot{a}_{em}}{\Delta t}, \tag{25}$$

Therefore, the redshift drift becomes:

$$\frac{dz}{dt_{abs}} = Q(t_{abs}, t_{em})\frac{t_{abs} - t_{em}}{a_{em}}\,. \tag{26}$$

The sign of $dz/dt_{abs}$ is thus determined by the sign of $Q(t_{abs}, t_{em})$.

Figure 3 illustrates the redshift as a function of cosmic time for galaxies at different comoving distances (units of $cH_0^{-1}$). The current redshifts are indicated by the points where the curves intersect the vertical line. Initially, the curves decrease due to deceleration in the early universe, but they later increase when the mean acceleration between emission and absorption becomes positive. Notably, galaxies with a current redshift around 2 have a minimum, indicating no redshift drift at the present time.

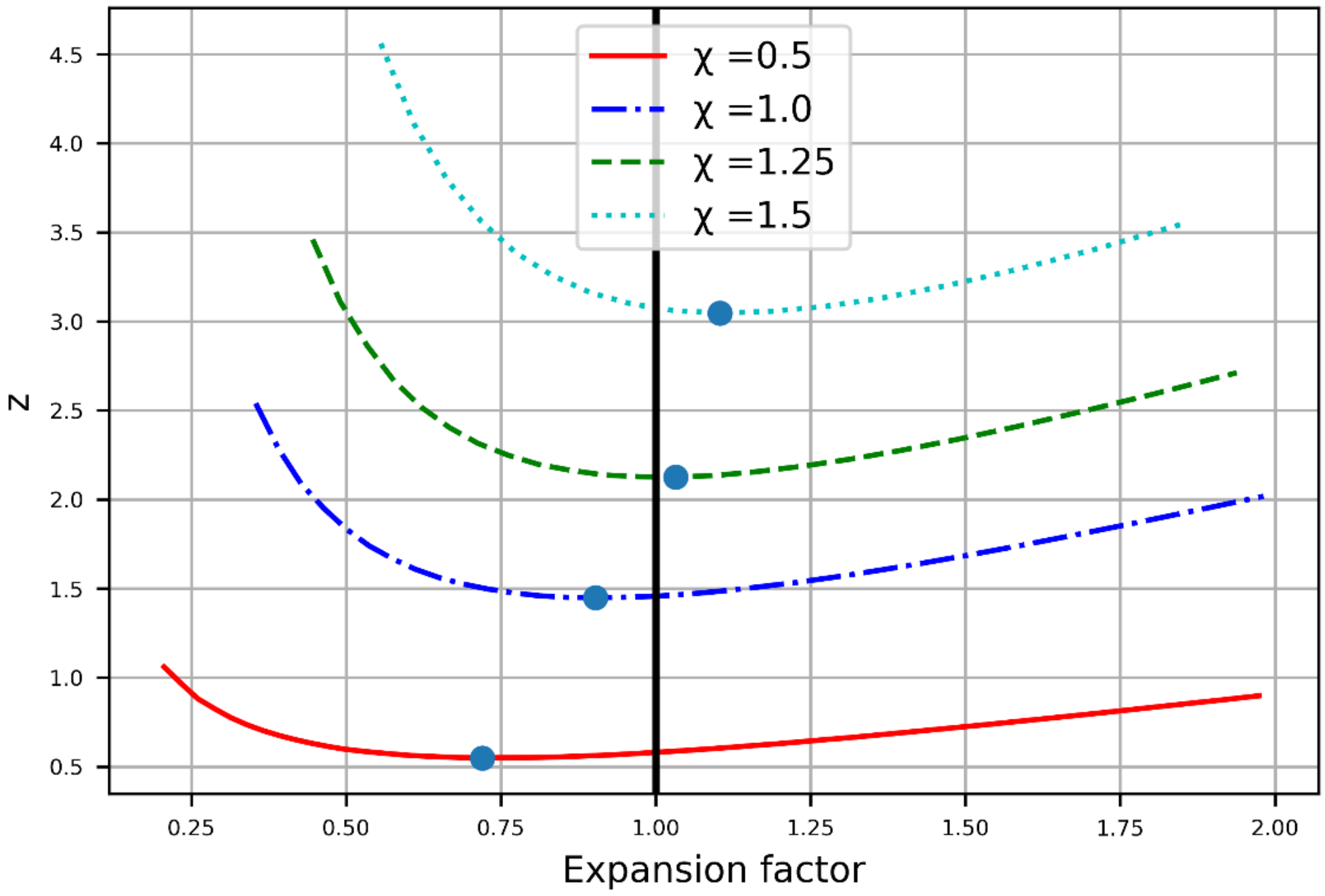

Figure 3. Redshifts as a function of cosmic time for galaxies at different comoving distances $\chi$ in units of $cH_0^{-1}$. The dots indicate the points where the redshift drift is zero.

The exact redshift where the redshift drift is zero at the present cosmological time is determined by combining Eq. 2 and 24

$$\left(1 + z_{no\ drift}\right) = H(z)/H_0 = \sqrt{(1 + z_{no\ drift})^3\Omega_m + 1 - \Omega_m}\,, \qquad (27)$$

which has the solution:

$$z_{no\ drift} = \frac{1 - 3\Omega_m + \sqrt{-3\Omega_m^2 + 2\Omega_m + 1}}{2\Omega_m}. \qquad (28)$$

For $\Omega_m = 0.3$ the result is $z_{no\ drift} = 2.09$. Measuring the redshift drift around $z \sim 2$ will not distinguish between the ΛCDM model and a coasting universe. Note that $z_{no\ drift}$ differs from the redshift where the acceleration is zero, which occurs at $z = 0.67$ according to Eq. 3, and also from the redshift where $\Omega_m = \Omega_\Lambda$, which is at $z = 0.33$ .
Fig 4. Shows the redshift drifts for galaxies with various redshifts at the present time.

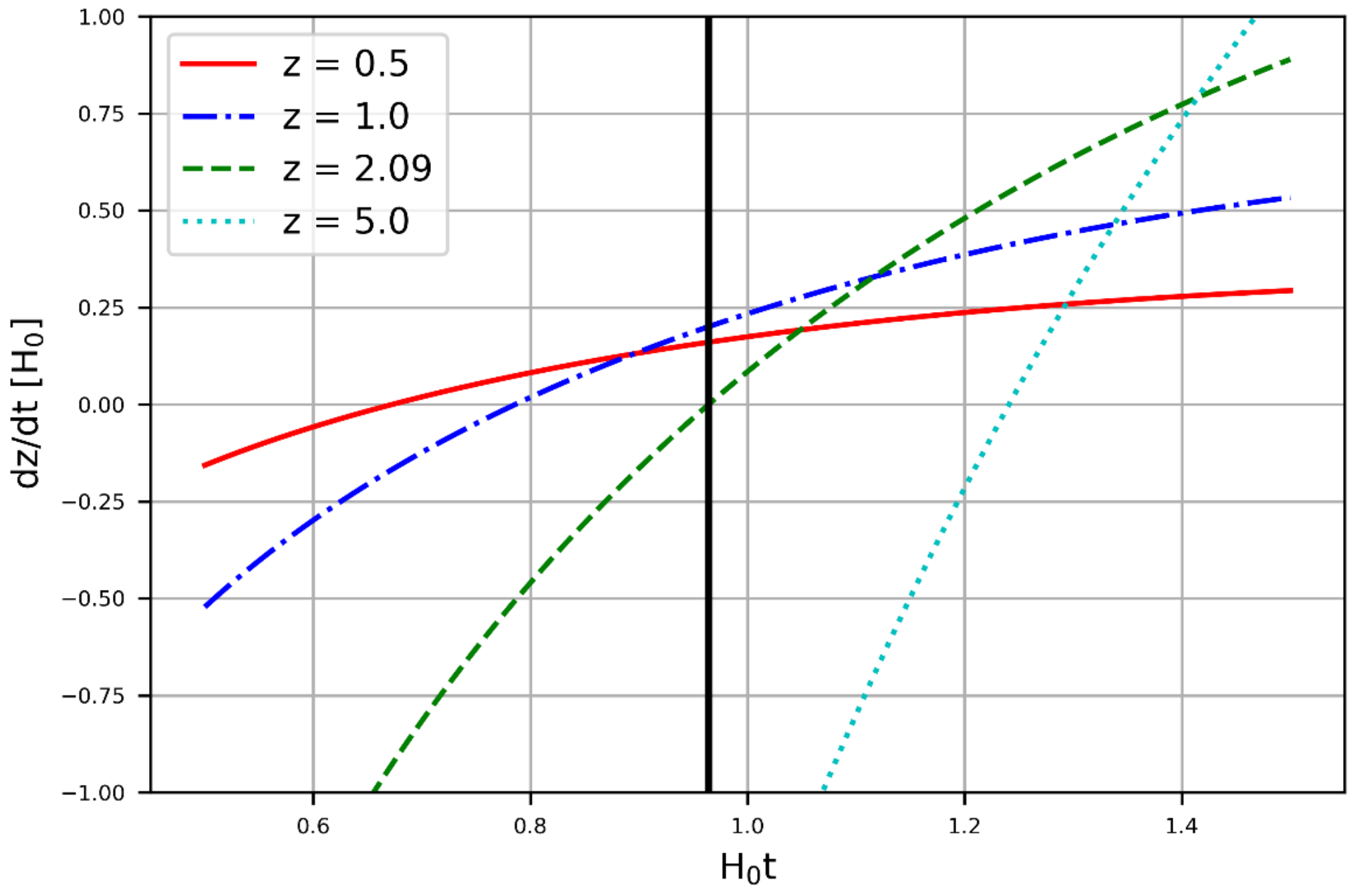

Figure 4. Change in redshift as a function of observation time for a flat $\Lambda CDM$ model with $\Omega_m = 0.3$. The thick vertical line indicates $H_0 t_0 = 0.964$

Finally we present a graph of the redshift drift as a function of $z_0$, the present redshift (see also [9] ).

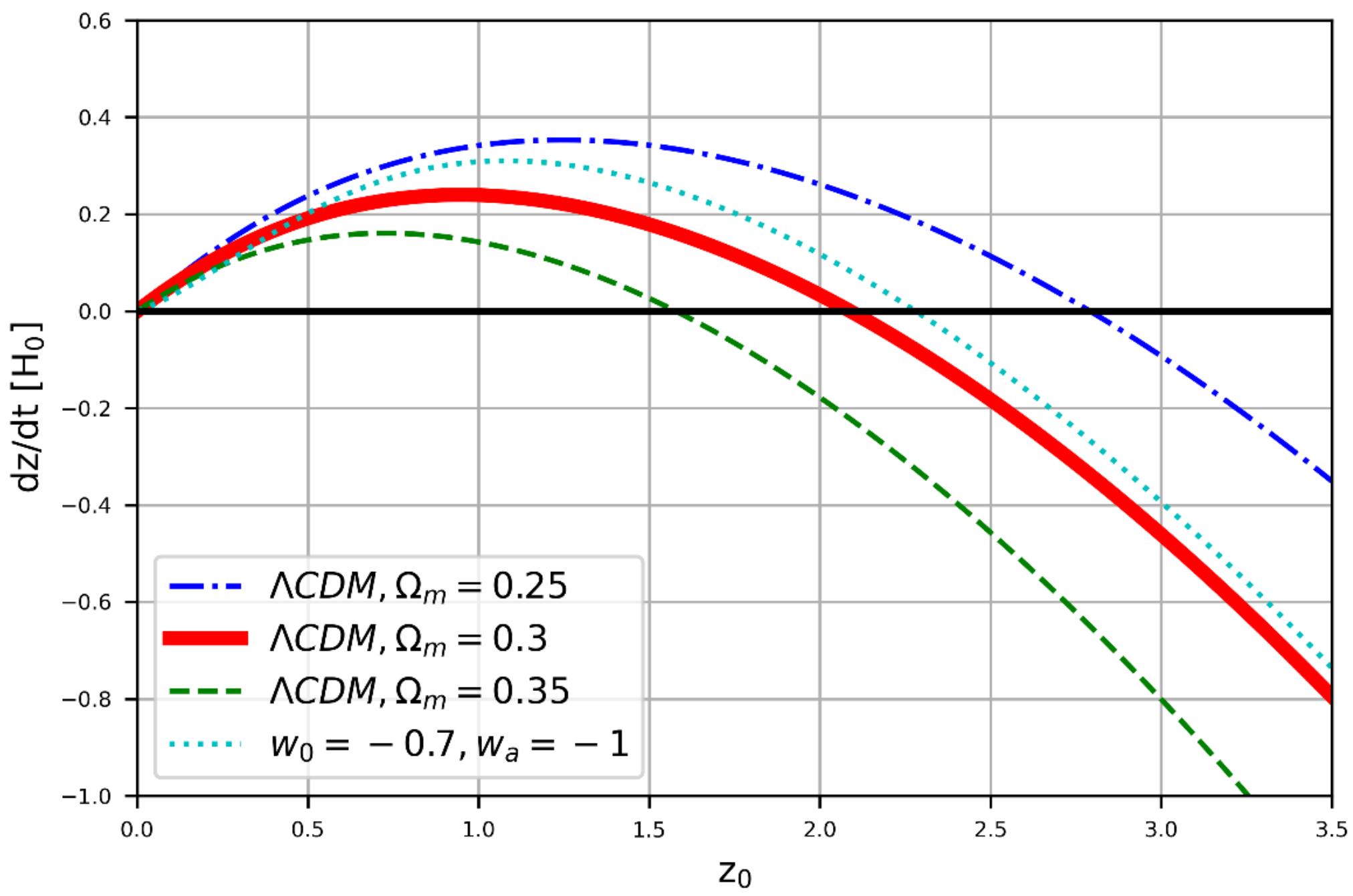

Figure.5 Redshift drift versus $z_0$ for flat $\Lambda CDM$ models with different $\Omega_m$ values and $w_\Lambda = -1$. The dotted line (just above the thick red line) is a model with $\Omega_m = 0.3$ and $w_{DE} = w_0 + w_a(1 - a)$.

In decelerating models without acceleration, such as Einstein-de Sitter model, the redshift drift will show only decreasing trends with $z_0$. In contrast models experiencing recent acceleration due to DE will exhibit a positive maximum in $dz/dt$ at $z{\sim}1$. The maximum redshift drift in the standard model is of order $0.3\ H_0 \approx\ 2\ 10^{-10}(10\ yr)^{-1}$ corresponding to a velocity change of $0.3\ cH_0 \approx 6\ cm\ s^{-1}(10\ yr)^{-1}$. This is a very small number, but detectable in future observations of large samples of galaxies and Lyman-alpha absorption lines in quasar spectra.

Individual galaxies tend to cluster in groups and clusters of galaxies. During this process they experience accelerations and develop peculiar velocities. Our Milky Way is part of the Virgo group with an infall velocity of approximately $300\ kms^{-1}$ to the group's centre and a velocity of $\sim 550\ kms^{-1}$ with respect to the CMB. [15] Over a Hubble time the mean acceleration from this infall is around $3-6\ 10^{-2} cm\ s^{-1}(10\ yr)^{-1}$.

Galaxies in rich virialized clusters may experience larger-than-average accelerations. A typical rich cluster might have a radius of $R_{cl}{\sim}2Mpc$ and velocity dispersion $\sigma{\sim}1000\ km\ s^{-1}$. In such clusters, galaxies experience a mean acceleration of approximately $\sigma^2/R_{cl} \approx 0.5\ cm\ s^{-1}(10\ yr)^{-1}$. Groups and clusters of galaxies broadly follow a mass – velocity dispersion relation in the form $M \propto \sigma^4$.[16] In combination with the virial theorem $\sigma^2 \propto M/R$ this would indicate that the acceleration ($\propto \sigma^2/R$) remains similar across a wide range of group and cluster sizes.

Still larger accelerations occur when galaxies fall towards each other and eventually collide. For example, the Milky Way and the Andromeda galaxy are expected to collide and merge in about $5\ 10^9 yr$. The velocity dispersion of the stars in both galaxies is on the order of $\sigma{\sim}200\ km\ s^{-1}$. As the galaxies draw near, the maximum acceleration occurs when their separation is roughly equal to their radius $R_g{\sim}10\ kpc$. The acceleration can be estimated as: $\sigma^2/R_g \approx 4\ cm\ s^{-1}(10\ yr)^{-1}$. This value is comparable to the cosmological redshift drift estimated earlier. However, the fraction of galaxies currently in the final merging stage – which lasts approximately $R_g/\sigma{\sim}10^7 yr$ - is expected to be small. Even if every galaxy undergoes one major merger in its lifetime, the fraction of galaxies presently merging is around $10^7 yr/t_0{\sim}10^{-3}$. Therefore, these mergers will not significantly affect the mean acceleration in large samples of galaxies.

The random accelerations that galaxies experience during the formation of groups and clusters of galaxies create a noisy background in the redshift drift measurements. However, the amplitude of this noise decreases with the inverse square of the number of galaxies in the sample. H.-R. Klockner et al. aim at observations of a billion HI galaxies, divided into redshift bins of $10^7$ galaxies per redshift interval, which would reduce the noise from peculiar motions by a factor of approximately $10^{3.5}$. [10] Liske et al. also estimated that the error in the redshift drift measurement of a large number of Ly$\alpha$ lines in quasar spectra due to peculiar motion is of order $10^{-3} cm\ s^{-1}(10\ yr)^{-1}$ [9] Therefore, redshift changes due to peculiar velocities are not expected to significantly impact results, as long as the sample sizes are sufficiently large.

## VII. DISCUSSION

We have discussed redshift drift using the concept of conformal time and emphasized the direct relationship between the redshift drift of galaxies (or the drift of Lyα lines in quasar spectra) and the mean acceleration of the universe during the interval between the emission and absorption of photons from those galaxies. By monitoring the mean redshift of large samples of galaxies over a 10-year period with an accuracy of $10^{-10}$, it would be possible to measure this redshift drift. In the near future, major telescopes like the Extremely Large Telescope (ELT) and the Square Kilometre Array (SKA) will begin observing the spectra of millions of galaxies and quasars in the redshift range of 0 to 5 with high precision. Several observing programs will utilize these next-generation telescopes to track redshift drifts in the spectra of galaxies and quasars, enabling the measurement of cosmic expansion in a direct, model-independent way. There is strong confidence that these observations will provide new insights into the evolution of our universe within the next 10–20 years.

## ACKNOWLEDGEMENTS

We would like to thank the Leiden Observatory for its hospitality.